\documentclass[10pt,conference]{IEEEtran}

\usepackage{graphicx}
\usepackage{placeins}
\usepackage{pbalance}
\usepackage{float}
\usepackage{booktabs}
\usepackage{multirow}
\usepackage{amsmath,amssymb}
\usepackage{threeparttable}

\begin{document}

\title{Modeling Failure Dynamics in Time-Constrained Authentication Systems: Evidence of a Success Cliff in USSD Workflows}

\author{
    \IEEEauthorblockN{Aklile Seyoum Mamo, Amanuel Kebede, Anny Christelle Irakoze, Jema Ndibwile$^*$}
    \IEEEauthorblockA{
        College of Engineering, Carnegie Mellon University Africa, Kigali, Rwanda\\
        Email: amamo@alumni.cmu.edu, akebede@alumni.cmu.edu, cirakoze@andrew.cmu.edu, jndibwil@andrew.cmu.edu\\
        *Corresponding author
    }
}

\maketitle

\begin{abstract}
Time-constrained interactive systems such as USSD (Unstructured Supplementary Service Data)-based financial services operate under strict session limits and sequential user interaction. While stronger authentication mechanisms improve security, they also increase interaction complexity and time burden, potentially reducing transaction completion. In this work, we model the failure dynamics of such systems and investigate how authentication complexity interacts with user response time and network round-trip time to influence session success rate. We propose and implement a simulation-based framework to investigate these failure dynamics and formally define a non-linear failure phenomenon, termed the \textit{Success Cliff}, where session success rates sharply decline beyond a critical complexity threshold. Through controlled experiments, we quantify the trade-off between security and usability and identify conditions under which secure authentication workflows become operationally unreliable.
\end{abstract}

\section{Introduction}

USSD-based systems enable critical services such as mobile money, balance inquiries, and bill payments. However, USSD interactions are constrained by short session durations (typically around 120 seconds), sequential menu navigation, and reliance on user input.

To enhance security, authentication mechanisms such as PIN verification, confirmation prompts, verification requiring delivery from external subsystems (e.g., SMS OTP) are commonly introduced. While these mechanisms strengthen security, they also increase interaction complexity and time requirements.

This creates a fundamental challenge: secure authentication workflows may become difficult to complete within the constraints of the system. For users navigating a multi-step authentication flow under an invisible session timer \cite{itu2016qos}, each additional prompt compounds cognitive load \cite{panjwani2010usably} and reduces the available time margin for error recovery, a dynamic that our results demonstrate can tip from manageable degradation into abrupt session failure. Despite this, existing research primarily focuses on improving authentication mechanisms rather than understanding when and why these systems fail in practice.

In this paper, we move beyond system design and instead model failure dynamics in time-constrained authentication systems. We identify the Success Cliff as the point at which session completion collapses non-linearly once authentication complexity exceeds a critical threshold, and show that out-of-band blocking 
delay, not step complexity alone, is the necessary condition for its emergence, with direct implications for authentication design in resource-constrained deployments.

\section{Related Work}

\subsection{Authentication in USSD and Mobile Financial Systems}

USSD-based financial systems are widely used to provide banking and mobile money services in regions with limited smartphone penetration. These systems rely on menu-driven interaction and session-based communication, requiring users to complete transactions within a fixed time window \cite{dayang2021ussd}.

Prior work has primarily focused on strengthening authentication protocols and identifying vulnerabilities in USSD communication. For example, Lamoyero and Fajana \cite{lamoyero2023ussd} analyze USSD banking systems and show that many platforms rely on weak or single-factor authentication, making them vulnerable to attacks.

Other studies propose improved authentication mechanisms, such as dynamic knowledge-based authentication models that generate user-specific challenge questions to enhance security \cite{njuguna2020ussd}. Similarly, research on mobile authentication systems highlights the increasing use of multi-factor authentication techniques to mitigate risks such as SIM swapping and social engineering \cite{kinai2020mfa}.

\subsection{Limitations of Existing Approaches}

Despite advances in authentication mechanisms, existing studies rarely evaluate how these systems perform under real-world interaction constraints. USSD systems operate under strict temporal and structural limitations, including short session durations, sequential menu navigation, and dependence on user response time.

In practice, each additional authentication step increases both cognitive and temporal burden. User interaction delays, network round-trip time, and input errors accumulate over the course of a session, increasing the likelihood of exceeding the system time limit or user abandonment.

However, most prior work evaluates authentication mechanisms in isolation, without considering the combined effect of these factors on session success rate. For instance, while Lamoyero and Fajana \cite{lamoyero2023ussd} highlight weaknesses in authentication design, they do not analyze how interaction complexity impacts usability or success rates.

Studies on mobile authentication often focus on improving security properties without modeling user interaction constraints or time-bound system behavior \cite{kinai2020mfa}. 

As a result, the operational reliability of secure authentication workflows in USSD systems remains under-explored, and system design has largely emphasized security guarantees while treating user interaction constraints as secondary considerations.

\subsection{Research Gap}

Existing research focuses primarily on the design and security of authentication mechanisms but does not systematically measure when and why these mechanisms fail under real-world constraints.

In particular, there is limited understanding of how authentication complexity interacts with user behavior and network conditions to influence session outcomes. Furthermore, prior studies do not capture potential non-linear failure behavior, where small increases in complexity may lead to disproportionate drops in transaction success. 

This gap motivates the need for a system-level analysis of failure dynamics in time-constrained authentication systems. In this work, we address this gap by modeling the interaction between authentication complexity, time accumulation, and user response time, and by identifying the conditions under which system performance collapses, referred to as the \textit{Success Cliff}.

\section{Problem Formulation}

We study USSD authentication workflows under strict time constraints to determine the point at which increasing authentication complexity becomes operationally non-viable, modeling the relationship between authentication complexity and session success to identify critical thresholds beyond which usability collapses. III-A–III-C establish the session model and core notation; III-D adapts the standard Keystroke-Level Model \cite{card1983psychology} to estimate user interaction time, while III-E–III-G model network round-trip time and blocking delay as stochastic processes calibrated to documented USSD conditions; III-H defines our complexity metric; and III-I–III-K introduce our primary modeling contributions, the abandonment models and the formal Success Cliff definition.

\subsection{System Model}
We model a USSD session as a sequence of $N$ interaction steps consisting of base navigation steps and authentication steps. Each step contributes to the total session time. 

\subsection{Variable Definition}
Core session variables are defined in Table \ref{tab:variables}.
\begin{table}[tb]
\centering
\caption{Variable Definitions for USSD Session Model}
\label{tab:variables}
\resizebox{\columnwidth}{!}{%
\begin{tabular}{lp{6cm}}
\hline
\textbf{Symbol} & \textbf{Description} \\ \hline
$k$ & Index of the current interaction step \\
$B$ & Number of base navigation steps \\
$A$ & Number of authentication steps \\
$N$ & Total number of steps ($N = A + B$) \\
$T_k$ & Cumulative session time elapsed after step $k$ \\
$t_k$ & Time elapsed at step $k$ \\
$U_k$ & User response time at step $k$ \\
$R_k$ & Network round-trip time at step $k$ \\
$H_k$ & Blocking delay at step $k$; nonzero only for steps 
        requiring delivery from a subsystem external to the active USSD session or out-of-band authentication (e.g.,\ SMS OTP) \\
$D_k$ & Retry delay incurred at step $k$ \\
$Z_k$ & Binary indicator of abandonment at step $k$ 
        ($Z_k \in \{0, 1\}$) \\
$P_{\text{abandon}}(k)$ & Probability of user abandonment at step $k$ \\
$\tau_s$ & Fixed session timeout threshold \\
$\tau_u$ & Fixed user response timeout threshold \\
$\tau_a$ & Fixed application timeout threshold \\ \hline
\end{tabular}%
}
\end{table}

\subsection{Session Time Model}

A session consists of a sequence of step attempts. Each step attempt $k$ incurs a total time cost defined as:

\begin{equation}
t_k = U_k + R_k + D_k + H_k .
\end{equation}

The components $U_k$, $R_k$, $D_k$, and $H_k$ are summarized here and are defined in detail in the following subsections.

The total session time is given by:

\begin{equation}
T_{\text{session}} = \sum_{k=1}^{K} t_k ,
\end{equation}

where $K$ is the total number of steps attempted, including retries.

\subsection{User Response Time Model}
User response time is modeled using a formulation inspired by the Keystroke-Level Model (KLM) \cite{card1983psychology}:

\begin{equation}
U_k = \left( t_{\text{read},k} + t_{\text{input},k} + t_{\text{context},k} \right).
\end{equation}

\subsection{Network Round Trip Time}

Network round-trip time $R_k$ represents the elapsed time between a user submitting a response and the next prompt being displayed. This includes network transmission, backend processing, and platform delays. It is modeled as a stochastic variable drawn from a Gamma distribution, with parameters chosen to produce mean latencies consistent with documented USSD response times under stable, congested, and degraded network conditions respectively \cite{dayang2021ussd, itu2017dfs}.

\subsection{Blocking Delay}

For authentication steps requiring delivery from external subsystems such as SMS OTP, the session incurs a blocking delay $H_k$ representing the 
time between the system dispatching the one-time code and the user 
receiving it. During this interval the USSD session timer continues to 
elapse, but neither user interaction nor system USSD processing occurs.

This delay is distinct from $R_k$ and $U_k$.

$H_k$ is modeled as a stochastic variable:

\begin{equation}
H_k = 
\begin{cases}
W_{\text{SMS}} & \text{if step } k \text{ is an SMS OTP step} \\
0 & \text{otherwise},
\end{cases}
\end{equation}

where $W_{\text{SMS}}$ is drawn from a predefined distribution 
reflecting SMS delivery latency. Based on SMS OTP delivery 
times under normal network conditions, we model:

\begin{equation}
W_{\text{SMS}} \sim U(5, 30) \text{ seconds}.
\end{equation}

\subsection{Retry Mechanism and Error Modeling}

When an error occurs, the user may retry the step, incurring an additional $U_k$, $R_k$, and $D_k$, with authentication steps carrying higher error probability than base steps due to increased interaction complexity.

\subsection{Complexity Definition}

Complexity is defined as a time-based proxy that reflects the cognitive and physical effort required by the user.
We model the complexity of an authentication step type $j$ as a relative measure based on expected user response time in comparison to a baseline authentication method:

\begin{equation}
C_j = \frac{\mathbb{E}[U_j]}{\mathbb{E}[U_{\text{baseline}}]},
\end{equation}

where: 

$U_{\text{baseline}}$ corresponds to a baseline authentication method (e.g., a 4-digit PIN) 
and 
$C_j$ is defined as a step-type property evaluated for the average user, in order to ensure user-independent comparability across authentication mechanisms. 

\subsection{Session Termination Conditions}
A session may terminate under the following conditions:

\begin{itemize}
    \item \textbf{Successful completion}
    \item \textbf{Session timeout:} if $T_{\text{session}} > \tau_{\text{s}}$
    \item \textbf{User response timeout:} if $U_k + H_k > \tau_{\text{u}}$
    \item \textbf{Application timeout:} if $R_k > \tau_{\text{a}}$
    \item \textbf{Abandonment}
\end{itemize}

\subsection{Abandonment Models}

We do not claim that abandonment is independent of time pressure. Instead, we model abandonment as a pre-timeout disengagement process and evaluate whether its behavior differs from timeout under controlled variations of error count normalized by the number of auth steps $A$ and network round trip time. If abandonment were merely a proxy for timeout, its prevalence would scale uniformly with elapsed time. Our experiments test whether this assumption holds.

These models are intended to provide transparent and conservative representations of pre-timeout disengagement rather than precise behavioral prediction. Consequently, we define three alternative abandonment models. At each step attempt $k$, the user may abandon the session with probability $P_{\text{abandon}}(k)$:

\paragraph{Model A: Time-Based}
\begin{equation}
P_{\text{abandon}}(k) = \min\left(a \cdot \frac{T_k}{\tau_s},\ p_{\max}\right).
\end{equation}

\paragraph{Model B: Event-Based}
\begin{equation}
P_{\text{abandon}}(k) = \min \left( b \cdot \frac{E_k}{A_k + 1},\ p_{\max} \right).
\end{equation}

where:
\begin{itemize}
    \item $E_k$ is the cumulative number of errors up to step $k$,
    \item $A_k$ is the cumulative number of attempts of authentication steps up to step $k$ (including retries).
\end{itemize}

\paragraph{Model C: Combined}
\begin{equation}
P_{\text{abandon}}(k) = \min\left(
c_1 \frac{T_k}{\tau_s} +
c_2 \frac{E_k}{A_k + 1},\ p_{max}
\right).
\end{equation}

These assumptions are intended to provide a tractable and interpretable model rather than an exact behavioral representation.

\subsection{Success Cliff}

Under fixed time constraints, failure does not accumulate along 
a single path: beyond a critical complexity threshold, small 
increases produce disproportionate collapse through concurrent 
timeout mechanisms rather than gradual degradation, a phenomenon 
we term the \textit{Success Cliff}. We formalize this as follows.

Let ${P_{success}(C)}$ denote the probability of successful session completion at complexity level C. We first define the consecutive drop in the success rate as:
\begin{equation}
\Delta(C) = P_{\text{success}}(C) - P_{\text{success}}(C+1), \quad C \geq 1.
\end{equation}
To capture whether this is an accelerated degradation, we define:
\begin{equation}
\Gamma(C) = \Delta(C) - \Delta(C-1), \quad C \geq 2.
\end{equation}
The Success Cliff is then defined as:
\begin{equation}
Q = \min\!\left\{C : \Delta(C) \geq \delta \text{ and } \Gamma(C) \geq \gamma\right\},\
\end{equation}
where:
\begin{itemize}
    \item ${C}$ is the authentication complexity level, determined by the complexity values $C_j$ of the authentication steps present in that session,
    \item ${Q}$ is the detected Success Cliff,
    \item ${\Delta (C)}$ is the drop in the success rate between consecutive complexity levels,
    \item ${\Gamma(C)}$ is the increase in degradation between consecutive drops,
    \item ${\delta}$ is the minimum drop threshold,
    \item ${\gamma}$ is the minimum acceleration threshold.
\end{itemize}

\section{Research Questions}

To systematically investigate the failure behavior of time-constrained authentication systems, we define the following research questions:

\begin{itemize}
\item \textbf{RQ1:} How does the complexity of authentication affect the session success rate under realistic delay conditions?
\item \textbf{RQ2:} At what point does increasing authentication complexity cause a sharp degradation in session success rate (Success Cliff)?
\item \textbf{RQ3:} How do authentication complexity and network round-trip time jointly influence system reliability?
\end{itemize}

\section{Methodology}

\subsection{Step Type Parameterization}

Step interaction times are derived using a KLM-inspired decomposition 
\cite{card1983psychology}, adapted for the USSD feature phone context. 
Each step is decomposed into three components: a cognitive processing 
time $t_{\text{read},k}$ corresponding to the M (mental preparation) 
operator, a physical input time $t_{\text{input},k}$ corresponding to 
the K (keystroke) operator, and a context-switching time 
$t_{\text{context},k}$ corresponding to the H (homing) operator. 
Standard KLM operator values are used: M = 1.35s, K = 0.28s per 
keypress, and H = 0.40s \cite{card1983psychology, kieras2001using}. 
These values are adapted to the feature phone numeric keypad context, 
where pointing operators are absent and interaction is limited to 
numeric input and menu selection.

The 4-digit PIN baseline interaction time is anchored to empirical 
measurements from Panjwani and Cutrell \cite{panjwani2010usably}, who 
report mean task completion times for plain PIN entry in a 
developing-world mobile banking context with 34 participants. Remaining 
authentication step times are KLM-derived estimates scaled to this 
empirical baseline.

The three base navigation steps advance the transaction without verifying identity: menu navigation selects a service from the numeric menu, amount enters the transaction value, and confirm approves the assembled transaction.

Error probabilities $p_{\text{error}}$ are estimated based on 
interaction complexity, following the principle established in the 
authentication usability literature that longer and more cognitively 
demanding input sequences incur higher error rates 
\cite{panjwani2010usably}.

\begin{table}[h!]
\centering
\caption{Base Step Type Parameters for Average User}
\label{tab:base_step_params}
\begin{tabular}{llcccccl}
\hline
\textbf{Step Type} & 
$t_{\text{read}}$ & 
$t_{\text{input}}$ & 
$t_{\text{context}}$ & 
$\bar{U}$ (s) & 
$C_j$ & 
$p_{\text{error}}$ \\ 
\hline
Menu navigation & 1.35 & 0.56 & 0.00 & 2.78 & --- & 0.02 \\
Amount          & 1.35 & 1.68 & 0.00 & 4.41 & --- & 0.02 \\
Confirm         & 1.35 & 0.28 & 0.00 & 2.37 & --- & 0.02 \\

\hline
\end{tabular}
\end{table}

\begin{table}[h!]
\centering
\caption{Authentication Step Type Parameters for Average User}
\label{tab:auth_step_params}
\begin{threeparttable}
\begin{tabular}{llcccccl}
\hline
\textbf{Step Type} & 
$t_{\text{read}}$ & 
$t_{\text{input}}$ & 
$t_{\text{context}}$ & 
$\bar{U}$ (s) & 
$C_j$ & 
$p_{\text{error}}$ \\ 
\hline
4-digit PIN          & 1.35 & 1.40 & 0.00 & 4.00 & 1.00 & 0.05 \\
6-digit PIN         & 1.35 & 1.96 & 0.00 & 4.80 & 1.20 & 0.08 \\
SMS OTP (6-digit)    & 1.35 & 1.96 & 2.15 & 7.92 & 1.98 & 0.12 \\
Challenge Response$^{\dagger}$  & 2.70 & 1.12 & 0.00 & 5.54 & 1.39 & 0.10 \\
\hline
\end{tabular}
\begin{tablenotes}
\small
\item$^{\dagger}$Models a typed challenge-response step requiring memory recall (e.g., last three digits of registered phone number).
\end{tablenotes}
\end{threeparttable}
\end{table}

\subsection{Parameterization}
\label{sec:parameterization}

Simulation parameters are defined in Table~\ref{tab:sim_params}. 
Where possible, values are anchored to documented operator 
thresholds or reported empirical ranges. Parameters lacking 
direct empirical support are set conservatively and are subject 
to the sensitivity analysis described in 
Section~\ref{sec:sensitivity}.

\begin{table}[!t]
\centering
\caption{Simulation Parameters and Justification}
\label{tab:sim_params}
\resizebox{\columnwidth}{!}{%
\begin{tabular}{lllp{3.8cm}}
\hline
\textbf{Parameter} & \textbf{Symbol} & \textbf{Value} & 
\textbf{Justification} \\ \hline

Session timeout & $\tau_s$ & 120s & 
Standard MNO limit for third-party USSD services \cite{itu2017dfs, 
dayang2021ussd} \\[3pt]

User response timeout & $\tau_u$ & 30s & 
Conservative operator-side threshold \cite{dayang2021ussd}. 
30s reflects a permissive upper bound. \\[3pt]

Application timeout & $\tau_a$ & 15s & 
Reflects documented backend response limits 
imposed by MNO's on 
application server response time \cite{dayang2021ussd} \\[3pt]

Simulation runs & --- & 50,000& 
Sufficient for stable estimation of 
session success rates at $\pm$0.44\% precision at 
95\% confidence \\[3pt]

Network RTT (low) & $R_k$ & $\text{Gamma}(2,\ 0.50)$s, mean $\approx 1$s & 
Reflects stable urban GSM conditions \cite{dayang2021ussd} \\[3pt]

Network RTT (medium) & $R_k$ & $\text{Gamma}(2,\ 1.25)$s, mean $\approx 2.5$s & 
Reflects intermittent connectivity typical 
of peri-urban or congested network 
conditions in African deployments \cite{itu2017dfs} \\[3pt]

Network RTT (high) & $R_k$ & $\text{Gamma}(2,\ 2.50)$s, mean $\approx 5$s & 
Reflects degraded or rural GSM conditions \\[3pt]

Max abandonment prob. & $p_{\max}$ & 0.3 & 
Conservative cap ensuring abandonment 
models remain bounded and do not 
dominate session outcomes at low 
complexity levels \\[3pt]

Model A coefficient & $a$ & 0.3 & 
Conservative value ensuring 
$P_{\text{abandon}}$ reaches $p_{\max}$ 
only near session timeout; 
set symmetrically with $p_{\max}$ \\[3pt]

Model B coefficient & $b$ & 0.4 & 
Slightly higher than Model A to reflect 
that event-driven abandonment is 
more sensitive to errors than 
passive time pressure \\[3pt]

Model C coefficients & $c_1, c_2$ & 
$0.2, 0.3$ & 
Weights assign primary burden to 
error accumulation ($c_2$), secondary 
burden to time pressure ($c_1$); 
sum kept below 1.0 to avoid 
$p_{\max}$ saturation at low 
complexity \\ \hline

\end{tabular}
}
\end{table}
\subsection{Sensitivity Analysis}
\label{sec:sensitivity}

Since error probabilities $p_{\text{error}}$ and several interaction time estimates are derived from KLM decomposition rather than direct empirical measurement in USSD contexts, we conduct a sensitivity analysis to assess whether key findings are robust to variation in these parameters.

For each experiment, simulations are repeated under three parameter regimes: a baseline configuration using the values defined in Table~\ref{tab:auth_step_params}, a low-error regime in which all $p_{\text{error}}$ values are reduced by 50\%, and a high-error regime in which all $p_{\text{error}}$ values are increased by 50\%.
A second analysis sweeps the $\gamma$ detection threshold across four values
($\gamma \in \{0.01, 0.02, 0.03, 0.04\}$) with $\delta$ fixed at $0.02$.
Since $\gamma$ has no external operational anchor, this sweep tests whether
the cliff detection outcome is an artifact of the chosen threshold value or
a stable structural property of the authentication flow.

\section{Experimental Design}
\label{sec:experimental_design}
To evaluate the impact of authentication complexity and network conditions on USSD session success, we design four controlled experiments. Each experiment varies one primary factor while holding all others constant, enabling causal interpretation of results.

{Authentication Complexity Configurations:}
We define four authentication complexity configurations corresponding to common USSD transaction scenarios, as shown in Table~\ref{tab:auth-configs}.
\begin{table}[tb]
\centering
\caption{Authentication complexity configurations.}
\label{tab:auth-configs}
\renewcommand{\arraystretch}{1.05}
\setlength{\tabcolsep}{3pt}
\begin{threeparttable}
\begin{tabular}{lll}
\hline
Level & Scenario & Workflow steps \\
\hline
C1 & Standard / Low complexity    & Mn, Am, P, Co \\
C2 & Sensitive / Moderate complexity  & Mn, Am, P, Mn, P, Co \\
C3 & Account recovery / High complexity & Mn, P, Mn, P, CR, Co \\
C4 & High-value / High complexity  & Mn, Am, P, Mn, P, OTP, Co \\
\hline
\end{tabular}
\begin{tablenotes}
\small
\item[$\dagger$] Mn=menu navigation, Am=amount, P=4-digit~PIN, Co=confirm, CR=challenge response, OTP=SMS OTP.
\end{tablenotes}
\end{threeparttable}
\end{table}

\subsection{Experiment 0: Baseline}
\begin{itemize}
    \item Complexity: C1,
    \item User behavior: average user,
    \item Network condition: medium round-trip time,
    \item Abandonment: all three models applied.
\end{itemize}
 
\subsection{Experiment 1: Authentication Complexity}
\begin{itemize}
    \item Variable: Complexity (C1, C2, C3, C4).
    \item Fixed: 
    \begin{itemize}
        \item Network condition: medium network round-trip time,
        \item User behavior: average user,
        \item Abandonment: all models applied.
    \end{itemize}
\end{itemize}

\subsection{Experiment 2: Network Round Trip}
\begin{itemize}
    \item Variable: Round-trip time (low, medium, high).
    \item Fixed: 
    \begin{itemize}
        \item Complexity: C1,
        \item User behavior: average user.
    \end{itemize}
\end{itemize}

\subsection{Experiment 3: Combined Conditions}
\begin{itemize}
    \item Variables: 
    \begin{itemize}
        \item Complexity: C1, C2, C3, C4,
        \item Round-trip time: low, medium, high.
    \end{itemize}
    \item Fixed: 
    \begin{itemize}
        \item User behavior: average user.
    \end{itemize}
\end{itemize}

\section{Evaluation Metrics}
We evaluate system performance using outcome based metrics derived directly from the session model.

The following metrics are used:
\begin{itemize}
    \item {Session Success Rate:}
 The proportion of sessions that successfully reach the final step of the workflow before any termination condition is triggered. 
    \item {Timeout Rate:}
 The proportion of sessions that terminate due to exceeding system-defined time constraints, including $\tau_s$, $\tau_u$, $\tau_a$. 
    \item {Abandonment Rate:}
 The proportion of sessions terminated voluntarily by the user prior to timeout, as determined by the abandonment models. 
    \item {Average Completion Time:}
 The mean total session time $T_{session}$ computed over successfully completed sessions only, capturing the effective interaction duration under each configuration. 
    \item {Input Error Rate:}
 The average number of input errors per session, derived from cumulative error counts $E_k$, reflecting the interaction difficulty associated with different authentication configurations.
\end{itemize}

\section{Results}

\subsection{Baseline Experiment Results (Experiment 0)}
To establish a reference point for the behavior of the system, we first simulate the standard low-complexity flow (C1) under medium network round-trip time with an average user. Across all three abandonment models, the baseline session success rate ranges from 98.30\% to 99.19\%. Session timeouts are negligible at 0.00\% and application timeouts account for at most 0.04\% of sessions. Abandonment accounts for the small residual failure rate, ranging from 0.78\% under Model A to 1.66\% under Model C, reflecting the models' differing sensitivity to time pressure and error accumulation, respectively. 

These results confirm that a standard single-PIN USSD authentication flow is operationally viable under typical network conditions, and establish a stable reference point for subsequent experiments.

We evaluate results against a 98\% session completion threshold, which we adopt as a conservative lower bound for operational viability. The ITU-T Focus Group on Digital Financial Services proposed USSD completion failure ratios of 0.1\%–1\% as target ranges for network-level KPIs, implying expected success rates of 99\%–99.9\% under normal operating conditions\cite{itu2016qos}. Our 98\% threshold deliberately sits below this range, representing the point at which failure rates become materially disruptive to users who depend on USSD as their primary financial channel. Results falling below this threshold are therefore interpreted as conditions under which an authentication workflow would be operationally non-viable even by conservative standards.

\subsection{Effect of Authentication Complexity Without Blocking Delay (Experiment 1)}

Across C1 through C4, session completion declines gradually from 98.48\%–99.17\% to 95.23\%–96.88\%, a total drop of approximately 2.3–3.3 percentage points. No Success Cliff is detected under any abandonment model or error regime. Consecutive-level drops remain within 1–1.5 percentage points throughout, and session timeouts stay negligible. Abandonment grows gradually from $\approx {0.8\%}$ at C1 to $\approx 3.7\%$ at C4, reflecting increasing interaction burden rather than abrupt disengagement. 

These results establish that authentication complexity alone even including the cognitive and input costs of multi-step verification produces predictable, manageable degradation. The system remains operationally viable across the full complexity range tested. The source of non-linear collapse must therefore lie in the delivery mechanism rather than step complexity itself. 
\begin{figure}[!htbp]
    \centering
    \includegraphics[width=0.7\linewidth]{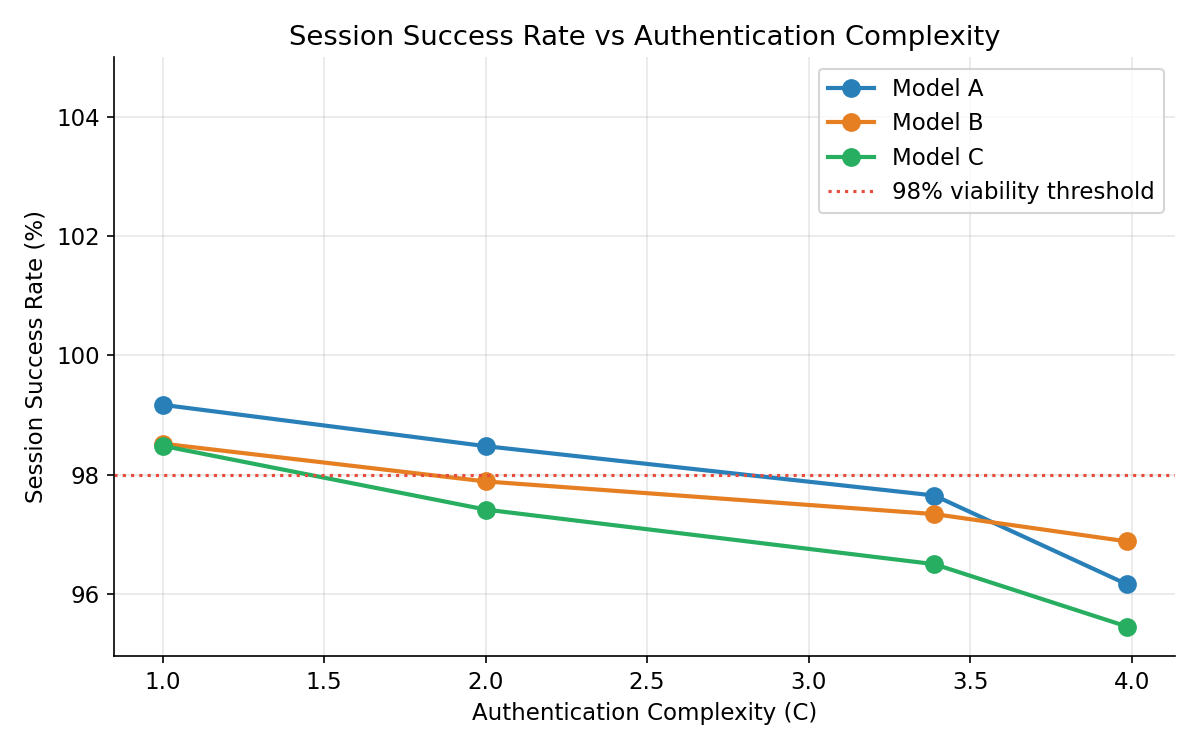}
    \caption{Authentication Complexity Vs. Success Rate without Blocking Delay}
    \label{fig:auth_success_without_block}
\end{figure}
 
\subsection{Effect of Network Round Trip Time (Experiment 2) }
Session success rates remain above 98\% under both low and medium latency, confirming that typical network conditions do not meaningfully threaten session viability for standard authentication flows. Under high latency, however, completion drops sharply to 91.64\%–92.20\%, falling below the 98\% operational viability threshold. This degradation is driven almost entirely by application timeouts, which rise to 6.59\%–7.05\% as individual round-trip times approach the application timeout threshold $\tau_a = 15$s. Abandonment remains largely stable across all three latency conditions, indicating that users do not disengage in response to latency alone.

No Success Cliff is detected under any latency condition. The degradation pattern is linear and attributable to a single mechanism, application timeout pressure, rather than the accelerated multi-factor collapse the cliff represents. These results confirm that high network latency is a meaningful independent threat to session viability, but that it operates differently from complexity-driven failure. Its effect is gradual and mechanistically distinct, which is important to interpret the combined conditions in the next experiment.

\begin{figure}[!htbp]
    \centering
    \includegraphics[width=0.7\linewidth]{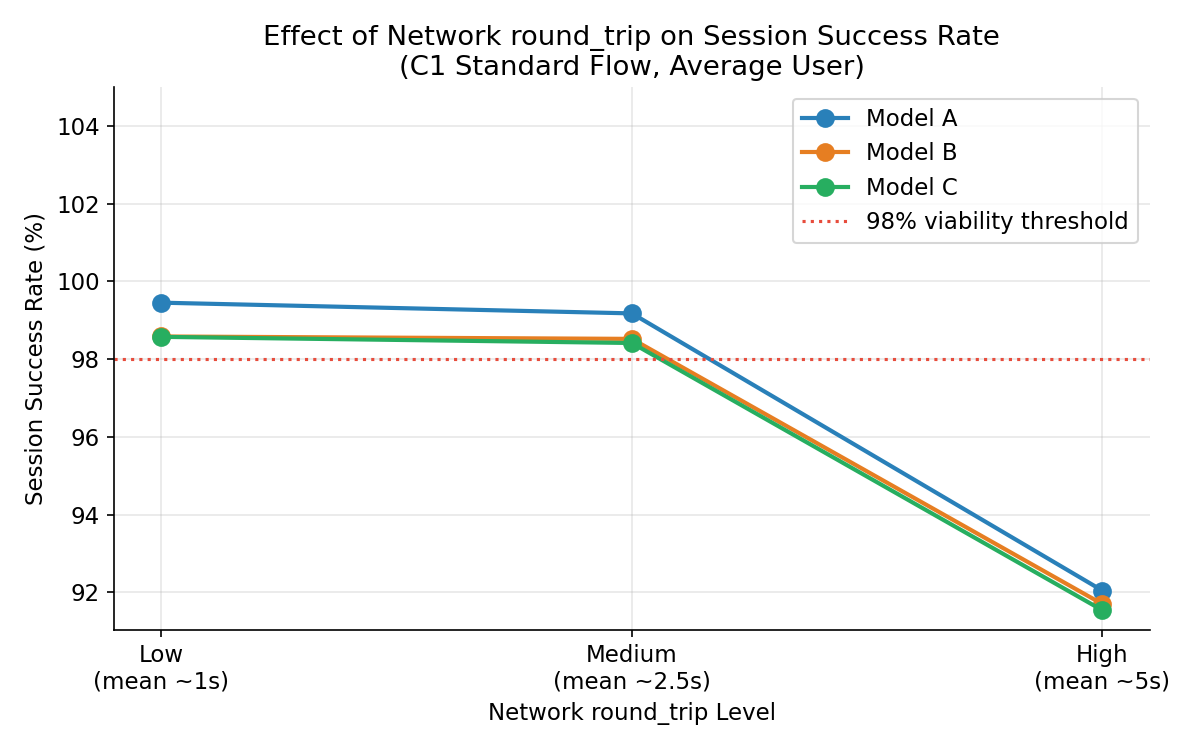}
        \caption{Network RTT vs Session Success Rate}
    \label{fig:network_success}
\end{figure}

 \subsection{Combined Conditions and the Role of Blocking Delay (Experiment 3)}

Experiment 3 jointly varies the complexity of authentication and the network round-trip time. We report results from both simulation conditions, with and without blocking delay, as the contrast between them identifies the mechanism driving non-linear failure.

\textbf{Without blocking delay}, the cliff emerges only under high network round-trip time at C4, where completion drops to 82.36\%–83.27\%. Under low and medium network round-trip time, completion remains above 95\% across all complexity levels with no cliff detected. Failure under high RTT is driven almost entirely by application timeouts (11.49\%–11.91\%), consistent with the pattern established in Experiment 2.

\textbf{With blocking delay}, a Success Cliff is detected across every network round-trip time condition tested, including low network round-trip time where network pressure alone produced no cliff. In each case, complexity level 3.39 (C3) represents the last viable operating point, with the sharp drop occurring at C4 (complexity 3.99) when the SMS OTP blocking delay is introduced. The C4 floor drops to 88.67\%–89.48\% under medium network round-trip time and collapses to 74.88\%–75.28\% under high network round-trip time. The blocking delay accounts for an additional 7–8 percentage points of failure on top of what network conditions alone produce.

\begin{figure}[!htbp]
    \centering
    \includegraphics[width=0.7\linewidth]{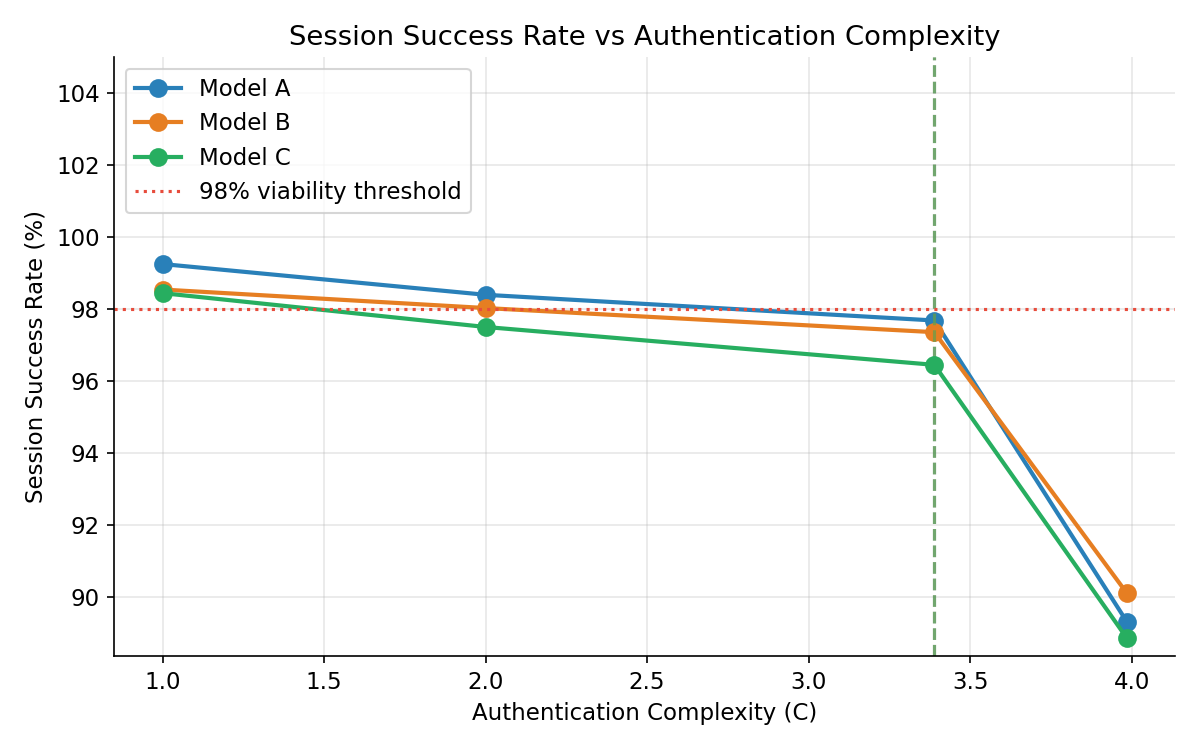}
    \caption{Authentication Complexity vs. Success Rate (With Blocking Delay)}
    \label{fig:complexity_success_graph_blocking}
\end{figure}

The failure composition reveals a mechanistic detail absent from the no-blocking-delay condition. Across all C4 blocking delay configurations, the four reported metrics: success, session timeout, application timeout, and abandonment, sum to approximately 95\% rather than 100\%, leaving a consistent residual of 4–5\% attributable to user response timeout. This residual is absent without the blocking delay. The explanation is architectural:  when $W_{\text{SMS}}$ draws a large value from $\mathcal{U}(5,30)$, the blocking delay consumes a significant portion of the per-step time budget before the user begins responding. Normal user response times that would otherwise complete successfully then push past $\tau_u = 30$s, triggering the user response timeout as a secondary failure mode. The blocking delay therefore generates two distinct failure pathways simultaneously: session timeout through cumulative time exhaustion, and user response timeout through per-step budget compression.

\begin{figure}[!htbp]
    \centering
    \includegraphics[width=0.7\linewidth]{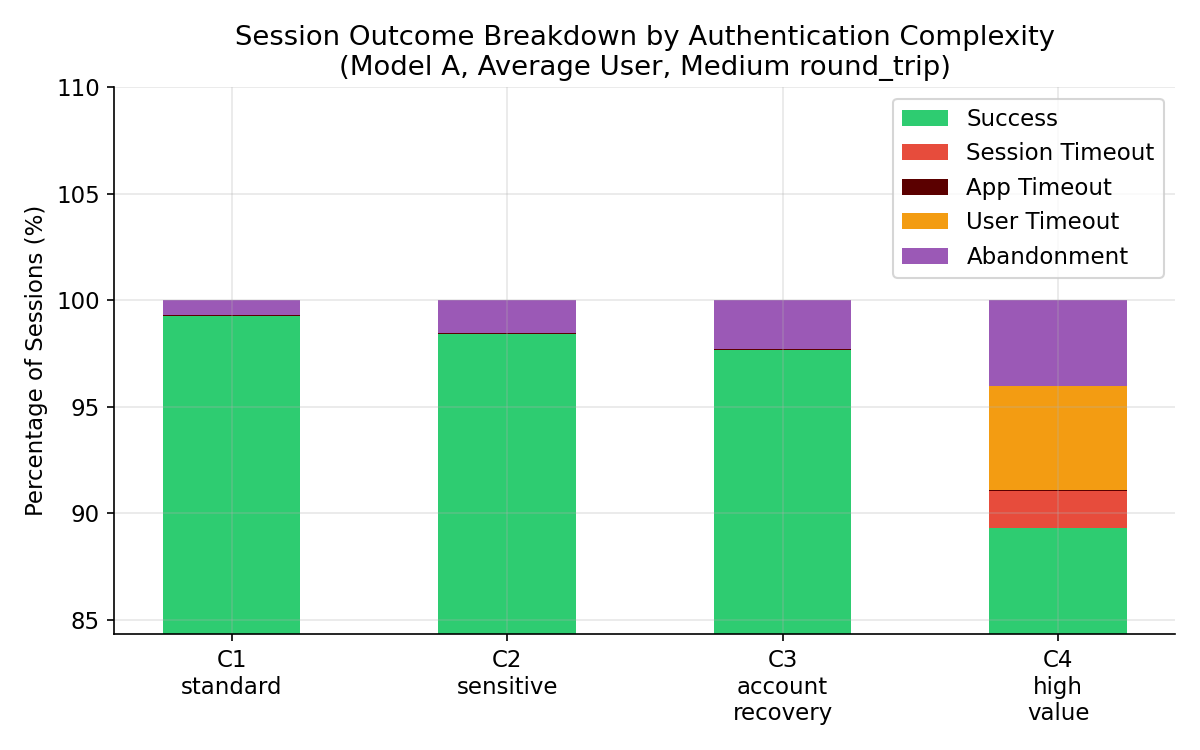}
    \caption{Failure Taxonomy with Blocking Delay}
    \label{fig:failure_tax_blocking}
\end{figure}

\begin{figure}[!htbp]
    \centering
    \includegraphics[width=0.7\linewidth]{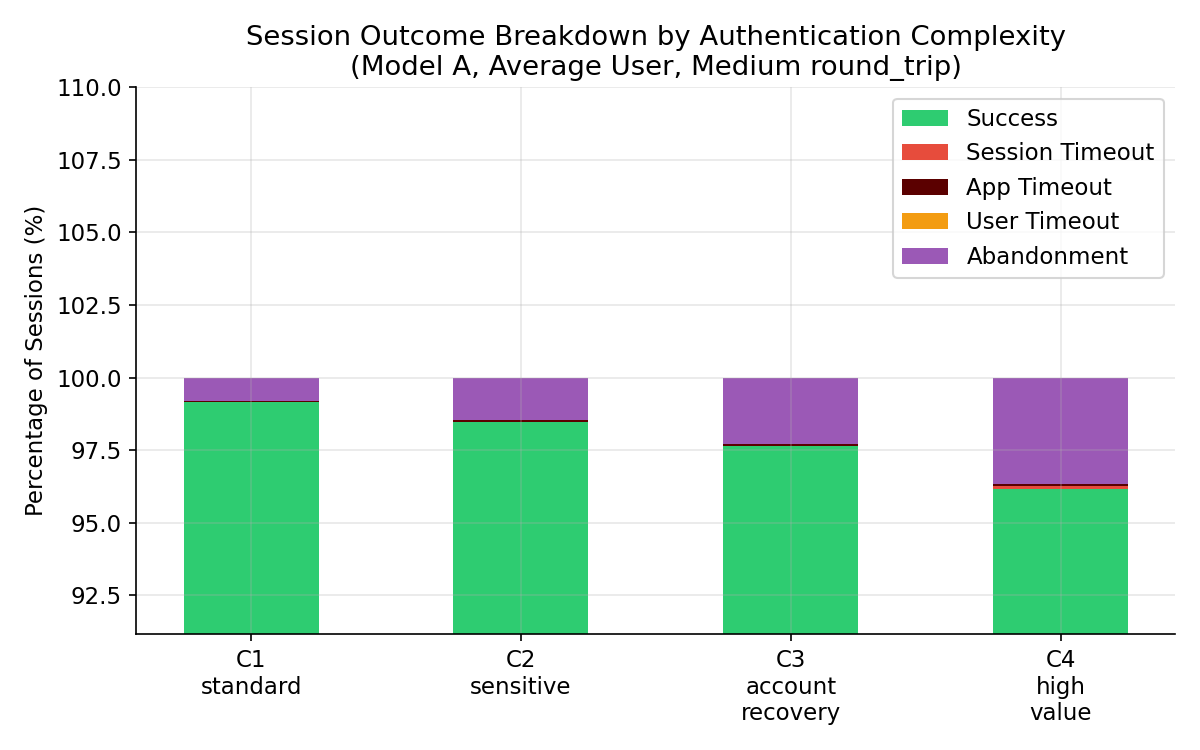}
    \caption{Failure Taxonomy without Blocking Delay}
    \label{fig:failure_tax_without_blocking}
\end{figure}

Together, these results demonstrate that the Success Cliff is not a product of complexity or latency acting independently. It emerges specifically when stochastic blocking delay interacts with the fixed session time budget, exhausting available time through multiple concurrent mechanisms, and producing failure rates that exceed the sum of their independent contributions.
\subsection{Effects of the Sensitivity Analysis}

\textbf{With blocking delay}, the cliff location is perfectly robust across both analyses. The transition from C3 to C4, where blocking delay is introduced via SMS OTP,  is detected as the cliff boundary under all three error regimes and across all 12 gamma threshold combinations (3 models × 4 values). The Q range across all combinations is 0.000. This confirms that within our experimental configurations, the cliff at the C3-to-C4 transition is insensitive to error probability assumptions and independent of the specific gamma threshold chosen, indicating it is driven by the structural introduction of blocking delay rather than by parameter choices or detection thresholds. 

\textbf{Without blocking delay}, neither analysis detects a cliff under baseline conditions. In the error probability analysis, a cliff emerges only under the high-error regime, when all $p_{\text{error}}$ values are artificially elevated by $ 50\%$ and is absent in both the baseline and low-error regimes. In the gamma threshold sweep, no cliff is detected across any of the 12 combinations. However, the absence of a cliff does not imply the system remains fully viable at high complexity. Without blocking delay, session success rates at C4 still fall to 95.23\%--96.88\% under baseline error assumptions, dropping below the $98\%$ operational viability threshold. The degradation is gradual and predictable rather than abrupt, but it is real, meaning that even in-band multi-step authentication imposes a meaningful operational cost on session completion as complexity accumulates.

\begin{table}[!htbp]
\centering
\caption{Cross-condition robustness of cliff detection across both sensitivity analyses. Robustness criterion: cliff location range $\leq \rho = 0.1$.}
\label{tab:sa-summary}

\scriptsize
\setlength{\tabcolsep}{4pt}
\renewcommand{\arraystretch}{1.1}

\begin{tabular}{llccc}
\toprule
\textbf{Condition} & \textbf{Analysis} & \textbf{Comb.} & \textbf{Cliff} & \textbf{Robust} \\
\midrule
\multirow{2}{*}{With blocking delay}
  & Error probability   & 9  & 9/9   & Yes \\
  & Gamma sweep         & 12 & 12/12 & Yes \\
\midrule
\multirow{2}{*}{Without blocking delay}
  & Error probability   & 9  & 1/9$^{\dagger}$ & No \\
  & Gamma sweep         & 12 & 0/12            & No \\
\bottomrule
\end{tabular}

\vspace{2pt}

{\footnotesize
$^{\dagger}$Cliff detected only under the high-error regime.}

\end{table}

These results sharpen the primary finding in two ways. First, the robustness of the C3-to-C4 cliff transition under blocking delay conditions confirms it is not a detection artifact of threshold selection or parameter uncertainty, but a structural consequence of introducing blocking delay into the session. Second, the contrast between conditions reveals two distinct failure regimes: a gradual, manageable degradation driven by in-band complexity accumulation, and a non-linear collapse triggered specifically by the stochastic delivery wait of out-of-band authentication. Without blocking delay, authentication complexity alone degrades session viability incrementally. With it, the same complexity configurations can produce abrupt, operationally unacceptable failure rates.

\section{Discussion}
\subsection{Blocking Delay as the Primary Architectural Threat}
The central finding is that blocking delay is the 
necessary condition for non-linear session failure. Complexity accumulation alone produces gradual, manageable degradation. A cliff emerges only under the most extreme network conditions when blocking delay is absent, confirming that it is the interaction between an uncontrollable delivery wait and a fixed session time budget that produces the abrupt collapse characterized by the Success Cliff.
\subsection{Two Distinct Failure Regimes}
The results reveal two qualitatively distinct failure regimes in time-constrained USSD authentication. The first is gradual degradation, where increasing authentication complexity incrementally consumes the session time budget, producing predictable declines in session success rate that remain detectable and bounded. The second is non-linear collapse, where the introduction of blocking delay creates a discrete, uncontrollable time cost that interacts with accumulated complexity and network latency to exhaust the session budget through multiple concurrent mechanisms simultaneously. These two regimes are not points on a continuum but structurally different failure modes, as evidenced by the sensitivity analysis: without blocking delay, a cliff is detected only when error rates are artificially elevated by $50\%$ beyond baseline estimates, suggesting the system sits near but below the cliff threshold under adverse parameter assumptions alone. With blocking delay, the cliff is invariant across all parameter variations tested. The transition between these regimes is therefore governed primarily by whether blocking delay is present, not by the magnitude of error or complexity assumptions.

It is also worth noting that even within the gradual degradation regime, sufficiently complex configurations are not without operational consequence: without blocking delay, session success rates at C4 fall to $ 95.23\% - 96.88\%$ under baseline assumptions, dropping below the $98\%$ viability threshold. This suggests that while complexity accumulation alone does not produce catastrophic failure, it is not operationally neutral either, and warrants attention independently of blocking delay considerations.

\subsection{Implications for Financial Inclusion}
The failure rates observed under combined blocking delay and high network latency, reaching as low as $74.88\%$ session completion, represent a material threat to operational reliability in exactly the deployment contexts where USSD is most critical. Rural deployments characteristically exhibit degraded network conditions and higher user response times relative to the urban averages most authentication benchmarks assume \cite{mwale2025zambia, okafor2022digital}, meaning the worst-case conditions modeled here are realistic rather than hypothetical for these populations.

SMS OTP based step-up authentication is not yet widely deployed for feature phone users in these contexts, so the cliff identified here does not represent a current exclusion risk. However, increasing fraud pressures and regulatory trends toward stronger authentication \cite{adewusi2025, mutabazi2025} suggest such mechanisms may be extended to these populations, making the failure dynamics identified here prospectively significant for financial inclusion.

\subsection{Limitations}
This study employs simulation rather than field measurement. 
KLM-derived interaction times use an empirical scaling factor 
anchored to PIN entry \cite{panjwani2010usably}, which may 
overestimate times for simpler steps and underestimate them 
for more cognitively demanding ones. Two mechanisms absent 
from the model would further increase failure rates in real 
deployments: per-step retry limits (e.g., MTN Uganda 
terminates sessions after three failed PIN attempts 
\cite{ali2020mfa}), and higher SMS delivery variability in 
rural or congested environments. The abandonment models are 
similarly conservative, though their consistency across all 
three formulations strengthens confidence in the structural 
finding even if absolute failure magnitudes are underestimated. 
Taken together, the failure dynamics reported here are likely 
conservative relative to real-world conditions.

\subsection{Future Work}
The most important next step is empirical validation. Field 
studies with real USSD users in contexts such as Rwanda, Uganda, 
Ghana, and Nigeria would ground key parameters user response 
times, error rates, blocking delay distributions, and abandonment 
behavior in observed rather than estimated data. Following 
validation, a natural extension is a design-time verification 
tool allowing USSD flow designers to evaluate whether a proposed 
authentication configuration is operationally viable for their 
target population before deployment.
\section{Conclusion}
This work models failure dynamics in time-constrained USSD authentication systems and identifies a non-linear collapse in session completion dubbed the ``Success Cliff'' that emerges specifically when authentication requires a stochastic delivery wait from a subsystem external to the active session. In-band complexity alone produces gradual and manageable degradation. It is the interaction between an uncontrollable blocking delay and a fixed session time budget that produces abrupt failure rates that fall well below operational viability thresholds. For populations that depend on USSD as their primary financial channel and for whom degraded network conditions are common, these findings suggest that authentication mechanisms intended to strengthen security may carry meaningful usability costs that warrant careful evaluation before deployment.

\end{document}